\documentclass{ws-procs9x6}
\def\d{\textrm{d}}

\begin{document}

\title{SMEARING DISTRIBUTIONS AND THEIR USE
IN FINANCIAL MARKETS}

\author{P. JIZBA$^*$ and H. KLEINERT$^\dag$}

\address{Institute for Theoretical Physics, Freie Universit\"at Berlin,\\
Arnimallee 14 D-14195 Berlin, Germany\\
$^*$E-mail: jizba@physik.fu-berlin.de\\
$^\dag$E-mail: kleinert@physik.fu-berlin.de}

%

\begin{abstract}
It is shown that superpositions of path integrals with arbitrary
Hamiltonians and different scaling parameters $v$ (``variances")
obey the Chapman-Kolmogorov relation for Markovian processes if and
only if the corresponding smearing distributions for $v$ have a
specific functional form. Ensuing ``smearing"  distributions
substantially simplify the coupled system of Fokker-Planck equations
for smeared and un-smeared conditional probabilities. Simple
application in financial models with stochastic volatility is
presented.
\end{abstract}


\bodymatter

\section{Introduction}\label{aba:sec1}

%
%

One often encounters in practical applications  probabilities
formulated as a superposition (or ``smearing'') of path integrals
(PI) of the form
\begin{eqnarray}
\mbox{\hspace{-2mm}}P(x_b,t_b|x_a,t_a) =  \!\int_{0}^{\infty} \d v \
\omega(v,t_{ba})\! \int_{x(t_a) = x_a}^{x(t_b) = x_b}{\mathcal{D}} x
{\mathcal{D}} p \ e^{\int_{t_a}^{t_b} \d \tau \left(i p\dot{x} - v
H(p,x) \right)} .\label{1.1}
\end{eqnarray}
Here $\omega(v,t_{ba})$ is some positive, continuous and
normalizable function  on ${\mathbb{R}}^+\!\times {\mathbb{R}}^+$
with $t_{ba}=t_b-t_a$ being the time difference. Examples of
(\ref{1.1}) can be found in financial markets~\cite{PI,jkh07}, in
polymer physics~\cite{PI,AKH}, in superstatistics~\cite{1,1b}, etc.

Whenever a smeared PI fulfills the Chapman-Kolmogorov equation (CKE)
for continuous Markovian processes, the Feynman-Kac formula
guarantees that such a superposition itself can be written as PI,
i.e.
\begin{eqnarray}
 P(x_b,t_b|x_a,t_a)\ = \
\int_{x(t_a) = x_a}^{x(t_b) = x_b}{\mathcal{D}} x {\mathcal{D}} p\
e^{\int_{t_a}^{t_b} \d \tau \left(i p\dot{x} - \bar
H(p,x) \right)}\, .
\end{eqnarray}
The new Hamiltonian $\bar H$ typically depends on first few momenta of
$\omega(v,t)$. 

Under normal circumstances, the smeared PI do not saturate CKE for
Markovian processes, in fact, ad hoc choices of smearing
distributions typically introduce memory into a dynamics. Our goal
here will be to isolate the general class of continuous smearing
distributions that conserve CKE. In Ref.~\cite{jk01} we have shown
that ensuing distributions have important applications in
evaluations of PI or in simplifications of the associated stochastic
differential equations. In this note we shall briefly present the
latter application. On the way we also mention a simple implication
in economical models with stochastic volatility.

\section{Most general class
of smearing distributions}\label{aba:sec2}

We look for $\omega(v,t)$ fulfilling CKE for any intermediate time $t_c$:
\begin{eqnarray}
P(x_b,t_b|x_a,t_a) \ = \ \int_{-\infty}^{\infty} d x \ \!
P(x_b,t_b|x,t_c) P(x,t_c|x_a,t_a)\, . \label{1.1a}
\end{eqnarray}
It can be shown~\cite{jk01} that Eq.(\ref{1.1a}) is fulfilled only
when the integral equation
\begin{eqnarray}
\int_{0}^{z}\! d z' \ \omega(z',t)\ \! a \ \!
\omega\!\!\left(\!a(z-z'),\frac{t}{a}\! \right)  = b\ \!
\omega\!\!\left(\!b z,\frac{t}{b}\! \right)\, , \label{1.1b}
\end{eqnarray}
($a,b \in {\mathbb{R}}^+$ and $1 + 1/a = 1/b$) holds.  By defining
the Laplace image function $\tilde{\omega}$  as
\begin{eqnarray}
\tilde{\omega}(\xi,t) = \int_0^{\infty} \d z \ e^{-\xi z}
\omega(z,t), \;\;\;\;\; \Re \xi >0\, ,
\end{eqnarray}
Eq.(\ref{1.1b}) can be equivalently formulated as the functional
equation for $\tilde{\omega}$
\begin{eqnarray}
\tilde{\omega}(\xi,t)\
\!\tilde{\omega}\!\left(\!\frac{\xi}{a},\frac{t}{a}\! \right) =
\tilde{\omega}\!\left(\frac{\xi}{b},\frac{t}{b}\right)\, .
\end{eqnarray}
Assumed normalizability and positivity of $\omega(v,t)$ implies
that smearing distributions are always Laplace transformable. After setting
$\alpha =1/a$ we get
\begin{eqnarray}
\tilde{\omega}(\xi,t)\
\!\tilde{\omega}\!\left(\alpha \xi, \alpha t\! \right) =
\tilde{\omega}\!\left(\xi + \alpha\xi, t +
\alpha t\!\right) \, . \label{2.3}
\end{eqnarray}
This equation can be solved by iterations. An explicit general solution
of Eq.(\ref{2.3}) was found in Ref.~\cite{jk01} and it reads
\begin{eqnarray}
\tilde{\omega}(\xi,t) \ = \ \left\{
                             \begin{array}{ll}
                               [G(\xi/t)]^t, & \hbox{when~~$t \neq 0\, ,$} \\
                               \kappa^{\xi}, & \hbox{when~~$t = 0\, .$}
                             \end{array}
                           \right.  \label{2.6}\end{eqnarray}
$G(x)$ is an arbitrary continuous function of $x$. Constant $\kappa$ is determined through the initial-time value of  $\omega(v,t)$. In particular,  $t \rightarrow 0$ solution (cf., Eq.(\ref{2.6})) gives
\begin{eqnarray}
\lim_{\tau\rightarrow 0} \omega(v,\tau) \ = \ \theta(v + \log \kappa)
\delta(v + \log \kappa) \ = \ \delta^+(v + \log \kappa)\,
.\label{1.9e}
\end{eqnarray}
Let us also notice that (\ref{2.3}) implies  $\tilde{\omega}(\zeta,
\tau)>0$  for all $\tau$ and $\zeta$ which gives $G(x)>0$  for all
$x$. This allows us to write
\begin{eqnarray}
[G(\zeta/\tau)]^{\tau} = e^{-F(\zeta/\tau)\tau}\, .\label{2.7}
\end{eqnarray}
Here $F(x)$ is some continuous function of $x$. Final $\omega(v,t)$
can be obtained via real Laplace's inverse transformation known as
Post's inversion formula~\cite{post}:
\begin{eqnarray}
\omega(v,t) = \lim_{k\rightarrow \infty} \frac{(-1)^k}{k!}
\left(\frac{k}{v}\right)^{\!\!k+1} \left.
\frac{\partial^k\tilde{\omega}(x,t)}{\partial x^k}\right|_{x= k/v}\,
.\label{2.9}
\end{eqnarray}
Real inverse transform (\ref{2.9}) is essential because the solution
of the functional equation (\ref{2.3}) was found only for real
variables in $\tilde{\omega}(\xi,t)$. In fact, complex functional
equations are notoriously difficult to solve.

We finally point out that the result (\ref{2.6}) is true also in the case when 
$v H(p,x) = v H_1(p,x) + H_2(p,x)$, such that $[H_1,H_2] =0$.

\section{Explicit form of $\bar{H}$}\label{aba:sec4}

To find $\bar{H}$ we use Post's formula (\ref{2.9}) which directly
gives
\begin{eqnarray}
&&P(x_b,t_b;x_a,t_a)\nonumber \\[1mm]
&&= \ -\lim_{k\rightarrow \infty} \int_{0}^{\infty}\d x \
\frac{(-x)^{k-1}}{(k-1)!}
\frac{\partial^k\tilde{\omega}(x,t)}{\partial x^k} \int_{x(t_a) =
x_a}^{x(t_b) = x_b} \!\!{\mathcal{D}} x {\mathcal{D}} p\
e^{\int_{t_a}^{t_b} \d \tau \ \! (i p\dot{x} -  k H/x)}
\nonumber \\[2mm]
&&= \ \int_{x(t_a) = x_a}^{x(t_b) = x_b} \!\!{\mathcal{D}} x
{\mathcal{D}} p \left[\int_{0}^{\infty} \d y \ \omega(y,t) \ e^{-
y\int_{t_a}^{t_b} \d \tau H }   \right] e^{i\int_{t_a}^{t_b} \d \tau
\
\! p\dot{x}}\nonumber \\[2mm]
&&= \  \int_{x(t_a) = x_a}^{x(t_b) = x_b} \!\!{\mathcal{D}} x
{\mathcal{D}} p  \ e^{\int_{t_a}^{t_b} \d \tau \ \! (i p\dot{x} -
F(H))}\, . \label{3.1}
\end{eqnarray}
In passing from second to third line we have used the asymptotic
expansion of the modified Bessel function $K_{k}(\sqrt{k} x)$ for
large $k$. On the last line we have utilized the definition of the
Laplace image.
%
Result (\ref{3.1}) thus allows to identify $\bar{H}$ with  $F(H)$.
Let us finally mention that the normalization
\begin{eqnarray}
1 = \int_{0}^{\infty} \d v \ \omega(v,t) = \tilde{\omega}(0,t)\label{3.3}
\end{eqnarray}
implies that $F(0) = 0$.
%
%

\section{Kramers-Moyal expansion for $\omega(v,t)$}\label{aba:sec5}

Let us now mention a simple application in stochastic processes. To
this end we notice that both $P(x_b,t_b|x_a,t_a)$ and
\begin{eqnarray}
P_v(x_b,t_b|x_a,t_a) \equiv  \int_{x(t_a) = x_a}^{x(t_b) = x_b}
\!\!\!{\mathcal{D}} x {\mathcal{D}} p\ e^{\int_{t_a}^{t_b} \d \tau
\left(i p\dot{x} - v H \right)}\! , \label{5.1}
\end{eqnarray}
fulfil CKE, so they can be alternatively evaluated by solving
Kramers-Moyal's (KM) equations~\cite{PI,vK}:
\begin{eqnarray}
&&\frac{\partial}{\partial t_b} P(x_b,t_b|x_a,t_a) =
{\mathbb{L}}_{KM}\ \! P(x_b,t_b|x_a,t_a)\, ,  \label{5.2a} \\
&&\frac{\partial}{\partial t_b} P_v(x_b,t_b|x_a,t_a) =
{\mathbb{L}}^v_{KM}\ \! P_v(x_b,t_b|x_a,t_a)\, ,   \label{5.2}
\end{eqnarray}
where the KM operator ${\mathbb{L}}_{KM}$ is
\begin{eqnarray}
{\mathbb{L}}_{KM} = \sum_{n=1}^{\infty}\left(-\frac{\partial
}{\partial x_b}\right)^{\! n} \ \!D^{(n)}(x_b,t_b) \, , \label{5.3}
\end{eqnarray}
and similarly for ${\mathbb{L}}^v_{KM}$. The coefficients $D^{(n)}$
and $D^{(n)}_v$ are defined through the corresponding short-time
transitional probabilities, so e.g., $D^{(n)}_v(x,t)$ is
\begin{eqnarray}
D^{(n)}_v(x,t) \ = \ \frac{1}{n!} \lim_{\tau\rightarrow 0}
\frac{1}{\tau}\int_{-\infty}^{\infty} \d y \ \! (y- x)^n P_v(y,t +
\tau|x,t)\, . \label{5.4bb}
\end{eqnarray}
%
Equations (\ref{5.2}) can be cast into an equivalent (and more
convenient) system of equations. For this we rewrite (\ref{1.1b}) as
\begin{eqnarray}
\omega(z,t)  =  \int_{0}^{\infty} \d z' \ \! \omega(z',t')
P^{\omega}(z,t|z',t')\, , \label{5.7}
\end{eqnarray}
with the conditional probability
\begin{eqnarray}
&&\mbox{\hspace{-14mm}}P^{\omega}(z,t|z',t')\ = \frac{t}{t-t'}\ \! \theta(t z - t' z') \
\!\omega\!\left(\frac{t z - t' z'}{t-t'},
t-t'\!\right)\!, \nonumber
\\
&&\mbox{\hspace{-14mm}}\int_{0}^{\infty} \d z \ \!
P^{\omega}(z,t|z',t') = 1\, , \;\;\;\;\;\;
\lim_{\tau \rightarrow
0}P^{\omega}(z,t+\tau|z',t) = \delta^+(z-z')\, . \label{5.8}
\end{eqnarray}
%
Eqs. (\ref{5.7})--(\ref{5.8})  ensure that the transition
probability $P^{\omega}(z,t|z',t')$ obeys CKE for a Markovian
process. Since the process is Markovian, we can define KM
coefficients $K^{(n)}$ in the usual way
as
\begin{eqnarray}
K^{(n)}(v,t)\ &=& \  \lim_{\tau\rightarrow 0} \frac{1}{n!\tau}
\int_{-\infty}^{\infty} \d x \ \! (x-v)^n
P^{\omega}(x,t+\tau|v,t)
\, . \label{5.9}
\end{eqnarray}
From ensuing CKE for a short-time transition probability one may
directly write down the  KM equation for $\omega(v,t_{ba})$
\begin{eqnarray}
\mbox{\hspace{-6mm}}\frac{\partial }{\partial t_{ba}} \ \! \omega(v,t_{ba}) =
{\mathbb{L}}^{(\omega)}_{KM} \ \! \omega(v,t_{ba})\, , \;\;\;
{\mathbb{L}}^{(\omega)}_{KM} = \sum_{n=1}^{\infty}
\left(-\frac{\partial}{\partial v}\right)^{\! n} K^{(n)}(v,t_{ba})\, .
\label{5.11aa}
\end{eqnarray}
%
%
%
In cases when both
(\ref{5.2}) and (\ref{5.11aa}) are naturally or artificially
truncated at $n=2$ one gets two coupled Fokker-Planck equations
\begin{eqnarray}
&&\mbox{\hspace{-9mm}}\frac{\partial }{\partial t} \ \! \omega(v,t)
= {\mathbb{L}}^{(\omega)}_{FP} \ \! \omega(v,t) , \;\;
\frac{\partial}{\partial t_b} P_v(x_b,t_b|x_a,t_a) =
{\mathbb{L}}^v_{FP}\ \! P_v(x_b,t_b|x_a,t_a), \label{5.11a}\\
&&\mbox{\hspace{9mm}}{\mathbb{L}}^{(\omega)}_{FP} =
{\mathbb{L}}^{(\omega)}_{KM}(n= 1,2)\, , \;\;\;\;
{\mathbb{L}}_{FP}^v = \mathbb{L}_{KM}^v(n=1,2)\, . \nonumber
\end{eqnarray}
On the level of sample paths the system (\ref{5.11a}) is represented
by two coupled It\={o}'s stochastic differential equations
\begin{eqnarray}
&&\d x_b = D^{(1)}_v(x_b,t_b) \ \!\d t_b +
\sqrt{2 D^{(2)}_v(x_b,t_b)} \ \! \d W_1\, , \nonumber \\
&&\d v = K^{(1)}(v,t_{ba}) \ \! \d t_{ba} + \sqrt{2 K^{(2)}(v,
t_{ba})} \ \! \d W_2\, . \label{5.13}
\end{eqnarray}
Here $W_1(t_b)$ and $W_2(t_{ba})$ are respective Wiener processes.
%

\section{Economical models
with stochastic volatility}\label{aba:sec6}

Note that in (\ref{5.13}) the dynamics of the variance $v$ is
explicitly separated from the dynamics of $x_b$. This is a desirable
starting point, for instance, in option pricing models~\cite{PI}. As
a simple illustration we discuss the stochastic volatility model
presented in Ref.~\cite{jkh07}. To this end take $G$ to be
\begin{eqnarray}
G(x) = \left(\!\frac{b}{x+ b}\!\right)^{\!\! c} \ \! , \;\;\;  b
\in {\mathbb{R}}^+; \;
c \in {\mathbb{R}}^+_0\, . \label{6.8f}
\end{eqnarray}
This gives
\begin{eqnarray}
\tilde{\omega}(\zeta,t) = \left(\frac{b t}{\zeta + b
t}\right)^{\!\!c t} \;\;\;\;\;\ \Rightarrow \;\;\;\;\;\;
\omega(v,t) =  \frac{(bt)^{ct} v^{ct -1}}{\Gamma(ct)} \ \! e^{-bt v}\, .
\label{6.11}
\end{eqnarray}
 $F(0)= 0$ as it should.
Distribution (\ref{6.11}) corresponds to the Gamma
distribution~\cite{feller,jkh07} $f_{bt, ct}(v)$.
The Hamiltonian $\bar{H}$ associated with (\ref{6.11}) reads
\begin{eqnarray}
\bar{H}(p,x) = \bar{v}b \ \! \log\!\left(\!\frac{H(p,x)}{b} + 1\right),
\label{6.13}
\end{eqnarray}
where $\bar{v} = c/b$ is the mean of $\omega(v,t)$. As $H$ we use
the Hamiltonian from Refs.~\cite{PI,jkh07} which has the form ${\bf
p}^2/2 + i{\bf p}(r/v -1/2)$, $r$ is a constant. This choice ensures
that $P_v(x_b,t_b|x_a,t_a)$ represents a riskfree martingale
distribution~\cite{PI}. Full discussion of this model without truncation is presented
in Ref.~\cite{jkh07}.

We consider here the truncated-level description that is epitomized
by It\={o}'s stochastic equations (\ref{5.13}). The corresponding
drift and diffusion coefficients $D_{v}^{(1)}$ and $D_{v}^{(2)}$ are
then (cf. Eq.(\ref{5.4bb}))
\begin{eqnarray}
\mbox{\hspace{-4mm}}D_{v}^{(1)}(x,t_{b}) \ = \  \left(r
-\frac{v}{2}\right),\;\;\;\;\;\;\;\;\;\;\;\;  D_{v}^{(2)}(x,t_{b}) \ = \ \frac{v}{2} .
\end{eqnarray}
For the coefficients $K^{(n)}$ an explicit computation gives
\begin{eqnarray}
K^{(1)}(v,t_{ba})  \ = \
\frac{1}{t_{ba}}
\left(\bar{v} - v\right) \, , \;\;\;\: K^{(2)}(v,t_{ba})\  = \
\frac{1}{t_{ba}^2} \frac{c}{2 b^2} \, . \label{5.14a}
\end{eqnarray}
Consequently, the It\={o}'s system takes the form
\begin{eqnarray}
\mbox{\hspace{-5mm}}\d x_{b}  =  \left(r -\frac{v}{2}\right)
\d t_{b} \ +
\sqrt{v} \ \! \d W_1 , \;\;\ \d v  =   \frac{1}{t_{ba}}\left(\bar{v} -
v\right)\ \! \d t_{ba} + \frac{1}{t_{ba}} \sqrt{
\frac{\bar{v}}{b}}
 \ \d W_2 . \label{6.36}
\end{eqnarray}
Let us now view $x_b$ as a logarithm of a stock price $S$, and $v$
and $r$ as the corresponding variance and drift. If, in addition, we
replace for large $t_{ab}$ the quantity  $\sqrt{\bar{v}}$ with
$\sqrt{{v}}$, the systems (\ref{6.36}) reduces to
\begin{eqnarray}
\d S \ = \ r S \ \!\d t_{b} \ +
\sqrt{v} S\ \! \d W_1\, , \;\;\; \d v \ =  \ \gamma\left(\bar{v} - v\right)\
\! \d t_{ba} + \varepsilon \sqrt{ {{v}}}
 \ \d W_2\, .
\label{6.36b}
\end{eqnarray}
The system of equations (\ref{6.36}) corresponds to Heston's
stochastic volatility model~\cite{PI,Heston1}, with mean-reversion
speed $\gamma = 1/t_{ba}$ and volatility of volatility $\varepsilon
= 1/(t_{ba}\sqrt{b})$. 


\section*{Acknowledgments}
This work has been partially supported by the Ministry of Education
of the Czech Republic under grant MSM 6840770039, and by the
Deutsche Forschungsgemeinschaft under grant Kl256/47.


\begin{thebibliography}{10}

\bibitem{PI}
H.~Kleinert, {\em Path {I}ntegrals in {Q}uantum {M}echanics, {S}tatistics,
  {P}olymer {P}hysics and {F}inancial {M}arkets}, 4th edn. (World Scientific,
  2004).

\bibitem{jkh07}
P.~Jizba,  H.~Kleinert, and P.~Haener, [arXiv:0708.3012] .

\bibitem{AKH}
A.~Kholodenko, {\em Ann. Phys.} {\bf 202}, p. 186 (1990).

\bibitem{1}
F.~Sattin, {\em Physica A} {\bf 338}, p. 437 (2004).

\bibitem{1b}
C.~Beck and E.~Cohen, {\em Physica A} {\bf 322}, p. 267 (2003).

\bibitem{post}
E.~Post, {\em Trans. Amer. Math. Soc.} {\bf 32}, p. 723 (1930).

\bibitem{jk01}
P.~Jizba and  H.~Kleinert, FU-Berlin preprint, to apper shortly


\bibitem{vK}
N.~V. Kampen, {\em Stochastic Processes in Physics and Chemistry}, 2nd edn.
  (North Holland, 1993).

\bibitem{feller}
W.~Feller, {\em An Introduction to Probability Theory and Its Applications,
  Vol. II}, 2nd edn. (John Wiley, 1966).

\bibitem{Heston1}
S.~Heston, {\em Trans. Amer. Math. Soc.} {\bf 6}, p. 327 (1993).

\end{thebibliography}

\end{document}